\documentclass[
prl,twocolumn,
amsmath,amssymb, aps,
]{revtex4}

\newcommand{\be}{\begin{equation}}
\newcommand{\ee}{\end{equation}} 
\newcommand{\bea}{\begin{eqnarray}} 
\usepackage{graphicx}
\newcommand{\eea}{\end{eqnarray}}

\usepackage{epsfig}
\usepackage{bm} 
\usepackage{helvet}
\usepackage{url} 
\usepackage{hyperref}
\usepackage[dvipsnames]{xcolor} 
\usepackage{dsfont}
\usepackage{amssymb}
\begin{document}

\title{Optimal Surfaces for Turbulent Circulation Statistics}

\author{L. Moriconi\footnote{moriconi@if.ufrj.br}}
\affiliation{Instituto de F\'\i sica, Universidade Federal do Rio de Janeiro, \\
C.P. 68528, 21945-970, Rio de Janeiro, RJ, Brazil and \\
Department of Mechanical and Aerospace Engineering, \\
New York University, New York, 11201, USA
}

\begin{abstract}
We put forward a novel formulation of the vortex gas model of turbulent circulation statistics to address the challenging case of nonplanar circulation contours. Relying upon a field-theoretical description, statistical moments of the circulation turn out to be functionally dependent on specific {\it{optimal surfaces}} bounded by the circulation loops. Circulation is modeled in the optimal curved spaces with the help of scalar vertex operators that represent the multifractal density fluctuations of Gaussian-correlated vortex structures. We show that minimal surfaces are optimal within the inertial range, but subdominant deviations are expected to become significant for contours with linear dimensions close to the Kolmogorov dissipation length.
As a case study, we demonstrate the model’s applicability through a Monte Carlo evaluation of the circulation probability distribution function for a nonplanar contour, which is in excellent agreement with results of extensive direct numerical simulations.
\end{abstract}

\maketitle

{\it{Introduction}}. The long-standing search for a deeper understanding of the complex interplay between convective transport and energy dissipation in turbulent flows has driven a remarkable variety of theoretical, computational, and experimental approaches \cite{Jorg-Sreeni}. 
Standing at the core of the ongoing debate about the fundamental aspects of turbulence, the mechanisms of energy transfer across scales (the turbulent cascade), small-scale dissipation, and inertial range intermittency have been extensively investigated in connection with the presence of strongly dissipative shear layers and vortex tubes \cite{sreeniRMP,ishihara_etal,jorg_etal,elsinga_etal, afonso_etal, afonso2_etal}. 

In this context, the circulation observable should naturally assume a special protagonism \cite{migdal2,cao_etal,benzi_etal,eyink}. Defined as the line integral of the velocity field along an oriented closed contour $C$, or, equivalently, as the vorticity flux through any oriented surface ${\cal{D}}$ bounded by $C$, viz.,
\be
\Gamma[C] \equiv \oint_C dx_i v_i(x) = \int_{\cal{D}} d^2 x \ {\bm{\hat n}}(x) \cdot {\bm{\omega}}(x) \ , \ 
\ee
velocity circulation works as a particularly convenient mathematical probe for examining vortical structures. 
Not fortuitously, shortly after the breakthrough discovery of circulation bifractality \cite{Iyer_etal}, a flurry of studies on circulation phenomenology has emerged, opening novel perspectives on both classical \cite{apol_etal,migdal20,bounded_measures,Iyer_etal2,mori_etal,mori_pereira,mori_etal2,Iyer_mori,muller_krst,duan_etal,mori_pereira2} and quantum turbulence \cite{muller_etal,polanco_etal,muller_etal2,muller_krst2}.

We are mainly interested in the intriguing relationship empirically found between the shape of the circulation probability distributions (cPDFs) and minimal surfaces in homogeneous and isotropic turbulence (HIT) \cite{Iyer_etal2,mori_PNAS}. Numerical evidence suggests, quite surprisingly, that the cPDFs of standardized circulation are uniquely determined by the minimal area enclosed by the circulation contour -- a phenomenon coined as the {\it{area rule}} of circulation statistics. A couple of interesting questions immediately come to mind: what kind of correlations between vortex structures would lead to the area rule? Are the circulation fluctuations actually well approximated by the total flux of localized vortex tubes? Considerable progress related to these issues has been achieved so far in the more specific setup of planar contours, along the lines of the vortex gas model (VGM) of circulation statistics \cite{apol_etal, bounded_measures,mori_etal,mori_pereira,mori_etal2,Iyer_mori,mori_pereira2}. In what follows, we propose a formulation of the VGM for nonplanar circulation contours, seeking to make explicit the role that minimal surfaces play in the statistical properties of turbulent circulation.

{\it{The Nonplanar Vortex Gas Model.}} To start, consider a closed contour $C$ in three-dimensional space that bounds an arbitrary surface given as the graph of $z=z(x)$, where $x=(x_1,x_2)$ and $z$ represent Cartesian coordinates. Let the surface's projection onto the $(x_1,x_2)$ plane be denoted by $\Sigma$. An example (to be further explored) is provided in Fig. 1.
\begin{figure}[h]
\hspace{0.0cm} \includegraphics[width=0.48\textwidth]{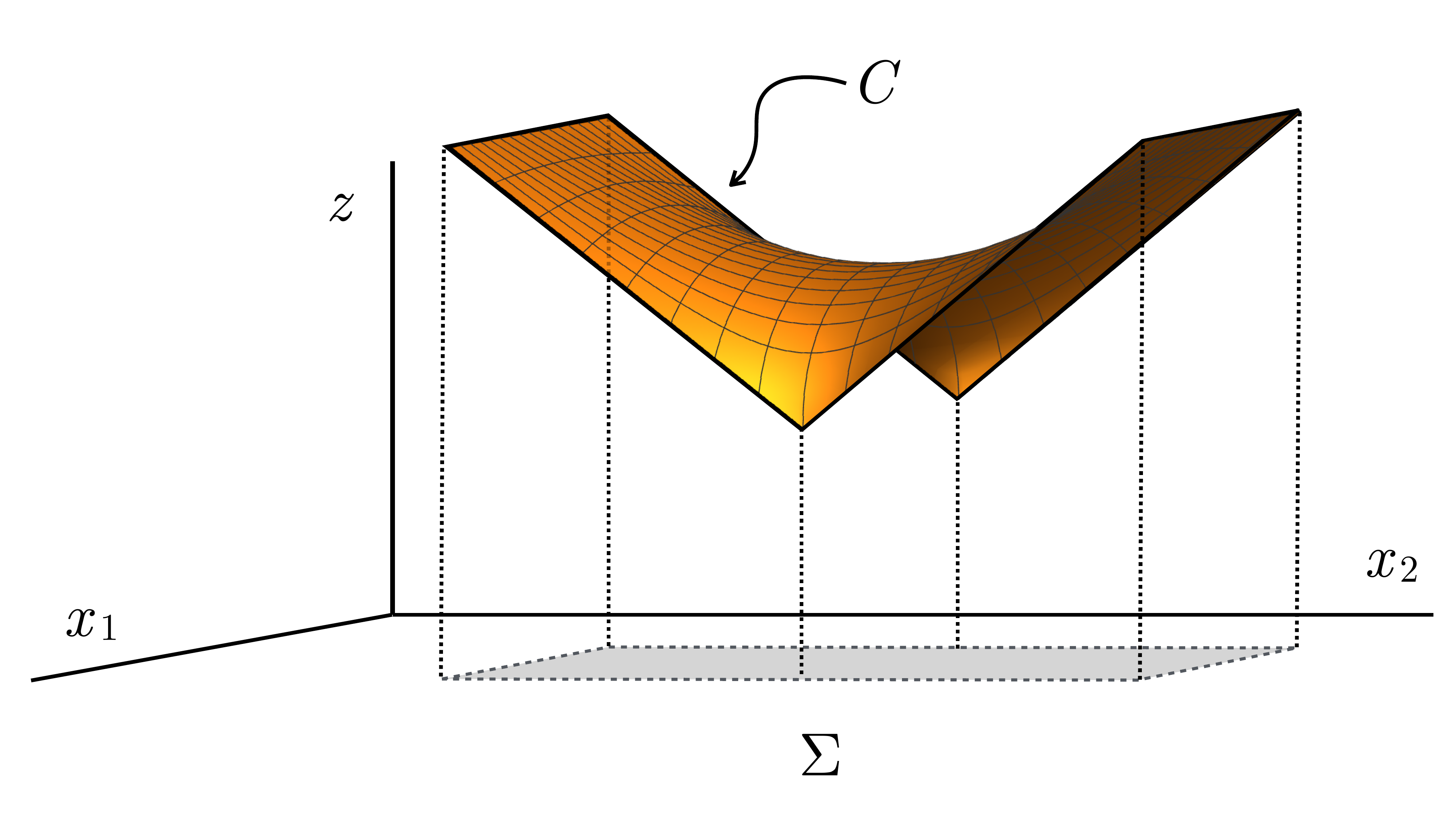}
\vspace{-0.3cm}
\caption{The polygonal circulation contour $C$ consists of six equal-length sides joined at orthogonal corners. Its projection onto the $(x_1, x_2)$ plane forms a rectangle that encloses the region $\Sigma$. Also displayed is the minimal surface bounded by $C$.}
\label{}
\end{figure}

Picking up an arbitrary orientation for $C$ and taking into account that vortex tubes form a dilute, locally polarized, and dense gas of structures \cite{mori_etal2}, we write the circulation around it as a functional of three fields:
\be
\Gamma[\phi,\tilde \Gamma,z] =
\int_\Sigma d^2 x \sqrt{g(x)} \xi(x,z) \tilde \Gamma(x,z) \ , \ \label{circ}
\ee
where (i) $d^2x \sqrt{g(x)}$, with $g(x) = \sum_i (\partial_i z)^2$, is the infinitesimal area element on the surface $z(x)$, (ii) $\xi(x,z)$ is the surface number density of the elementary vortices (vortex tubes) that cross the surface $z(x)$ at position $(x,z)$, and $\tilde \Gamma(x,z)$ is proportional to the circulation carried by them in a small neighborhood of the same surface position \cite{comment}.

In the VGM, the density field $\xi(x,z)$ is taken to be the same in probability law as $\sqrt{\epsilon(x,z)}$, where $\epsilon(x,z)$ is the energy dissipation rate at position $(x,z)$. This hypothesis, accurately validated from detailed analyses of DNS data \cite{mori_pereira, mori_pereira2}, renders the theory of Gaussian multiplicative chaos (GMC) a useful modeling tool \cite{GMC}. When applied to turbulence, the GMC approach yields a field-theoretical generalization of the Obukhov-Kolmogorov lognormal model of turbulent intermittency (OK62) \cite{O62,K62,pereira_etal}. Borrowing from the language of conformal field theory \cite{francesco_etal}, this amounts to saying that the energy dissipation field is modeled in the GMC formulation as the scalar vertex operator
\be
\epsilon(x,z) = \epsilon_0 \exp \left [ \gamma \phi(x,z) -\gamma^2 \langle \phi^2 \rangle/2 \right ] \ , \ \label{epsilon}
\ee
where $\phi(x,z)$ is a log-correlated Gaussian random field,
\bea
&&\langle \phi(x,z) \phi(x',z') \rangle = \nonumber \\
&&-\frac{1}{4 \pi} \ln \left ( \frac{(x-x')^2 + (z-z')^2+ \eta^2}{L^2} \right )
  \ , \ \label{phi-phi}
\eea
with $L$ and $\eta$ being, respectively, the integral and the Kolmogorov dissipation length scales of the flow, and {\hbox{$\gamma \equiv \sqrt{2 \pi \mu}$}}, where $\mu = 0.17 \pm 0.01$ is the usual intermittency exponent for the power law decay of energy dissipation correlations \cite{tang_etal}. All that means that we are led, from (\ref{epsilon}) and (\ref{phi-phi}), to $\epsilon_0 = \langle \epsilon(x,z) \rangle$ and
\be
\langle \epsilon(x,z) \epsilon(x',z') \rangle 
\approx \epsilon_0^2 \left [ \frac{(x-x')^2 + (z-z')^2+ \eta^2}{L^2} \right ]^{- \frac{\mu}{2}} \ , \
\ee
for {\hbox{$\eta \ll \sqrt{(x-x')^2 + (z-z')^2} \ll L$}}
(the inertial range scales).

Going much beyond the reach of the OK62 model, the GMC description of energy dissipation fluctuations provides a straightforward way to evaluate, from (\ref{epsilon}) and (\ref{phi-phi}), arbitrary multi-point correlation functions of $\epsilon(x,z)$. Furthermore, it follows from (\ref{epsilon}) that fields like $[\epsilon(x,z)]^q$ have scaling laws similar to the ones of the dissipation field itself, once $\mu$ is replaced by $\mu q^2$. The case $q=1/2$ corresponds to the density field $\xi(x,z)$. We introduce, with a convenient normalization,
\be
\xi (x,z) = \eta^{-2} \exp \left [ \gamma \phi(x,z)/2  \right ] \ . \  \label{xi}
\ee

The elementary circulation field $\tilde \Gamma(x,z)$, completely independent of $\xi(x,z)$, is, on its turn, described in the VGM as a zero-mean, self-similar Gaussian random field, with scaling dimension $[\tilde \Gamma ] \equiv \alpha = 2 - \mu/4 - \zeta_2 \approx 1.26$ \cite{mori_pereira2}, where $\zeta_2 \approx 0.7$ is the scaling exponent of second order velocity structure functions in HIT \cite{frisch}. Working with circulation units where the variance of $\tilde \Gamma$ is unit and considering that in the nonplanar setting the circulation of a vortex tube changes sign when the local surface orientation ${\bm{\hat n}}$ flips to 
${-\bm{\hat n}}$, the simplest expression for the correlator of $\tilde \Gamma(x,z)$ takes the form
\bea
&& \hspace{-1.0cm} \Delta(x,x') \equiv \langle \tilde \Gamma(x,z) \tilde  \Gamma(x',z') \rangle  = \nonumber \\
&& \hspace{-1.0cm} = \eta^\alpha \frac{ {\bm{\hat n}}(x,z) \cdot {\bm{\hat n}}(x', z')}{[(x-x')^2 + (z-z')^2 + \eta^2]^\frac{\alpha}{2}} \ . \ \label{Delta}
\eea
We trivially note that $\langle \tilde \Gamma^2 \rangle = \Delta(x,x) = 1$, as already mentioned, and that the nonplanar correlator (\ref{Delta}) reduces to the planar VGM results when ${\bm{\hat n}}(x,z) \cdot {\bm{\hat n}}(x', z') = 1$.

The remaining modeling ingredient of (\ref{circ}) concerns the specific choice of the surface $z=z(x)$. If the circulation were computed through an exact evaluation of the vorticity flux, any arbitrarily chosen surface would suffice, owing to the solenoidal nature of the velocity field.
Previous studies of the VGM have established that (\ref{circ}) yields faithful circulation statistics when planar surfaces are employed. In earlier works, this choice was adopted as an implicit modeling assumption; here, by contrast, the selection of an optimal surface demands explicit justification and deeper analysis.

In principle, there is no \textit{a priori} reason to expect the existence of a unique optimal surface. 
Considering a large statistical ensemble of flow configurations, and assuming that for each ensemble element 
the circulation is best approximated by the use of (\ref{circ}) through the choice of a particular surface 
\(z_j = z_j(x)\), with its associated vortex density distribution 
\(\xi_j = \exp[\gamma \phi_j(x,z)/2]\) 
and elementary circulation field 
\(\tilde{\Gamma}_j = \tilde{\Gamma}_j(x,z)\), the characteristic function of the random circulation is formally written as the expectation value
\be
\Phi(\lambda) =  \lim_{N \rightarrow \infty} \frac{1}{N} \sum_{j=1}^N \exp \{ - i \lambda \Gamma[\phi_j,\tilde \Gamma_j, z_j ] \}  \ . \ \label{cf}
\ee
Resorting to the independence of the elementary circulation field, we partially average (\ref{cf}) over the Gaussian fluctuations of $\tilde \Gamma$ to get
\be
\Phi(\lambda) =  \lim_{N \rightarrow \infty} \frac{1}{N} \sum_{j=1}^N \exp \left [ -  \frac{1}{2} \lambda^2 \langle \Gamma_j^2 \rangle_{\tilde \Gamma} \label{cf2}
\right ]  \ , \
\ee
where 
\bea
&&\langle \Gamma_j^2 \rangle_{\tilde \Gamma} = \int_\Sigma d^2 x \int_\Sigma d^2 x'  \times \nonumber \\
&&\times \sqrt{g_j(x)} \sqrt{g_j(x)} \xi(x,z_j) \xi(x',z_j') \Delta(x,x') \ . \ \label{Gamma2}
\eea
The summands in
(\ref{cf2}) correspond to different realizations of the scalar and surface fields, $\phi(x,z)$ and $z(x)$, respectively. Invoking the joint probability density functional $\rho[\phi,z]$, Eq. (\ref{cf2}) can now be recast as the following functional integral,
\be
\Phi(\lambda) = \int D[z] D[\phi] \rho[\phi,z]
\exp \left [ -  \frac{1}{2} \lambda^2 \langle \Gamma^2 \rangle_{\tilde \Gamma}
\right ] \ . \ \label{cf_pi}
\ee
Although the exact form of $\rho[\phi,z]$ is unknown, we assume it to be a formal, yet unnormalized, distribution over $\phi$ (with $z(x)$ held fixed) that reproduces the two-point correlation function given in Eq.~(\ref{phi-phi}). 
The Gaussian structure of $\rho[\phi, z]$ is, for this reason, putatively defined as
\be
\rho[\phi,z] = C \exp\{ - S[\phi,z] \} \ , \ \label{jointpdf}
\ee
where $C$ is a normalization constant and
\bea
&&S[\phi,z] = \frac{1}{2}\int_\Sigma d^2 x \int_\Sigma d^2 x' \times \nonumber \\ 
&&\times \sqrt{g(x)}\sqrt{g(x')}\phi(x) \phi(x') K(x,x'|z) \ , \ \label{actionS}
\eea
is a quadratic action given in terms of a nonlocal kernel $K(x,x'|z)$. Above (and from now on), we replace $\phi(x,z)$ by $\phi(x)$ to simplify the notation, having in mind that $z=z(x)$.

One might regard the use of covariant integration measures in Eq. (\ref{actionS}) as unnecessary, given that the correlation function (\ref{phi-phi}), and hence $K(x,x'|z)$, is not invariant under diffeomorphisms of the $(x_1,x_2)$ coordinates. The procedure, however, is well-founded: short-distance scalar correlations depend solely on the geodesic distance between points on the surface and are therefore invariant under coordinate diffeomorphisms. As it will become evident shortly, it is precisely this covariant ultraviolet behavior of the vortex density fluctuations that endows minimal surfaces with a central role in the description of circulation statistics.

The characteristic function (\ref{cf_pi}) becomes, after elementary manipulations,
\be
\Phi(\lambda) = \int D[z] \exp  \{ -H[z,\lambda]  \} \ , \ \label{phiH}
\ee
where
\be
H[z,\lambda]  \equiv F[z] - \ln C - \ln \left \langle 
\exp \left [ -  \frac{1}{2} \lambda^2 \langle \Gamma^2 \rangle_{\tilde \Gamma}
\right ] \right  \rangle_\phi \ , \  \label{H}
\ee
with
\be
F[z] \equiv - \ln \int D[\phi] \exp \{ - S[\phi,z] \} \ , \ \label{f_energy}
\ee
and
\bea
&&\left \langle 
\exp \left [ -  \frac{1}{2} \lambda^2 \langle \Gamma^2 \rangle_{\tilde \Gamma} \right ] \right  \rangle_\phi \equiv \exp \{ F[z] \} \times \nonumber \\
&&\times \int D[\phi] \exp \{ - S[\phi,z] \}  \left [ -  \frac{1}{2} \lambda^2 \langle \Gamma^2 \rangle_{\tilde \Gamma}
\right ] \ . \ 
\eea

It follows, from (\ref{actionS}) and (\ref{f_energy}), that up to some unimportant constant shift,
\be
F[z] = \frac{1}{2} \ln \det [K]  = - \frac{1}{2}\int_{\eta^2} \frac{dt}{t} {\hbox{Tr}} \{\exp [ -t K ] \} \ . \ \label{FzDet}
\ee
In the second equality above, we have applied the standard ultraviolet regularized heat-kernel, for a consistent evaluation of the determinant (see Sec. S1 of the supplemental material for a discussion). It turns out that at leading order,
\be
F[z] = \frac{1}{8 \pi} \frac{1}{\eta^2} \int_\Sigma d^2x \sqrt{g(x)} = \frac{1}{8 \pi} \frac{A[z(x)]}{\eta^2} \ , \ \label{Fz}
\ee
where $A[z(x)]$ is the area spanned by the surface bounded by the circulation contour. 

In view of (\ref{Fz}), the functional integrand in Eq.~(\ref{phiH}) contains a factor sharply peaked at the minimal surface value. To determine the optimal surface in the steepest descent evaluation of (\ref{phiH}), we must examine the saddle-point equation,
\be
\frac{\delta}{\delta z} H[z,\lambda] =
 \frac{1}{8 \pi \eta^2} \frac{\delta}{\delta z} A[z(x)] 
+ O(\lambda^2) = 0 \ . \ \label{sp}
\ee
One may write the solution of (\ref{sp}) as the series expansion
\be
z_\lambda(x) = z_0(x) + \eta^2 \sum_{n=0}^\infty \sum_{m=1}^\infty \eta^{2n} \lambda^{2m} z_{n,m}(x) \ , \ \label{expansion}
\ee
where $z_0(x)$ represents a minimal surface. Combining (\ref{phiH}) and (\ref{expansion}), the statistical moments of the circulation can be effectively evaluated as derivatives, at $\lambda=0$, of the saddle-point characteristic function
\be
\Phi(\lambda) = \exp \{ -H[z_\lambda,\lambda] + H[z_0,0] \} \ . \
\ee
However, as it follows from (\ref{expansion}), in the zero viscosity limit ($\eta \rightarrow 0$, with fixed contour) we have $z_\lambda (x) \rightarrow z_0(x)$, which
yields 
\be
\Phi(\lambda) = \exp \{ -H[z_0,\lambda] + H[z_0,0] \} \ . \
\ee
This observation underpins the analysis presented below.

{\it{Circulation Probability Distributions.}} Fourier transforming Eq. (\ref{cf2}), one then finds that an arbitrary cPDF $\rho(\Gamma)$ can be expressed as the expectation value of centered Gaussian distributions $\rho_\sigma(\Gamma)$ over an ensemble of variances $\sigma^2 = \sigma^2[\xi]$ given by (\ref{Gamma2}), with fixed minimal surface $z_0(x)$ and random vortex densities $\xi(x,z_0)$,
\be
\rho(\Gamma) = \langle \rho_\sigma(\Gamma) \rangle_\sigma \ . \ \label{rhoGamma}
\ee
Following the numerical investigation of HIT and the parameter settings of Iyer {\it{et al.}} at the Taylor Reynolds number $R_\lambda = 1300$ \cite{Iyer_etal2}, we discuss the circulation statistics for the polygonal contour $C$ shown in Fig.~1, {\it{via}} Eq. (\ref{rhoGamma}). Working in units of length for which $\eta = 1$, the integral length scale of the flow is $L = 2276$, while the line segments of $C$ have length $R = 150$.

An ensemble of $N=10^4$ random realizations of the scalar field $\phi(x)$ over a square lattice with spatial resolution $\eta$ is produced from a Gaussian functional probability measure that gives the two-point correlation function (\ref{phi-phi}). The numerical procedure is detailed in Sec. S2 of the supplemental material. Each realization of $\phi(x)$ is used to generate a sample of the surface vortex density (\ref{xi}) and thus one for the fluctuating variance (\ref{Gamma2}) in the circulation-dependent expectation value (\ref{rhoGamma}).

The modeled and the numerical cPDFs of Ref.~\cite{Iyer_etal2} are compared in Fig.~2. We show, in Fig.~2a, the cPDFs for the nonplanar contour depicted in Fig.~1, and for a planar square contour (of side $R' = 1.334R$) that has the same area as the minimal surface. The circulations in the plots of Fig.~2a are normalized by the standard deviations $\sigma_{np}$ of the nonplanar cPDFs.

A key observation that highlights the role of minimal surface geometry in turbulent circulation statistics is that the accurate modeling of the numerical cPDFs of Fig.~2a hinges on the orientational dependence of the $\tilde \Gamma$-$\tilde \Gamma$ correlations in Eq.~(\ref{Delta}). Had it not been for the ${\bm{\hat n}}(x,z) \cdot {\bm{\hat n}}(x', z') $ factor in (\ref{Delta}), the matches with numerical cPDFs would have been lost.

\begin{figure}[t]
\hspace{-0.2cm} \includegraphics[width=0.45\textwidth]
{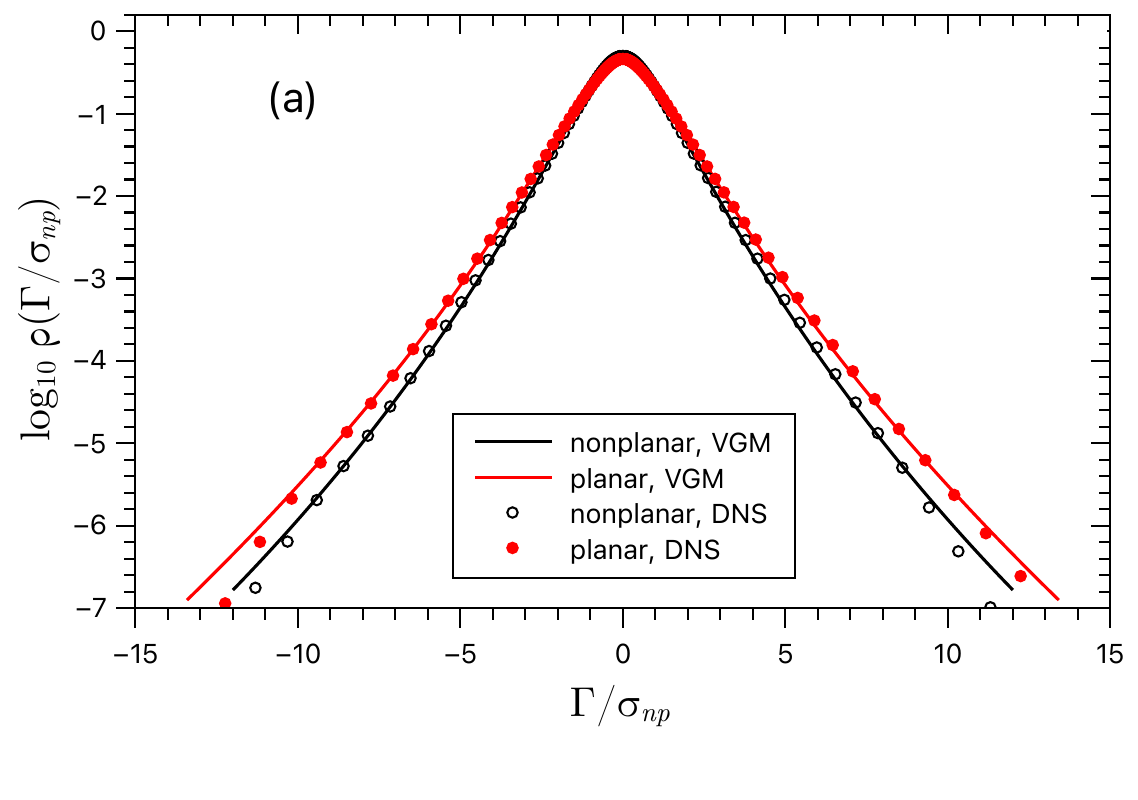}
\hspace{-0.5cm} \includegraphics[width=0.45\textwidth]{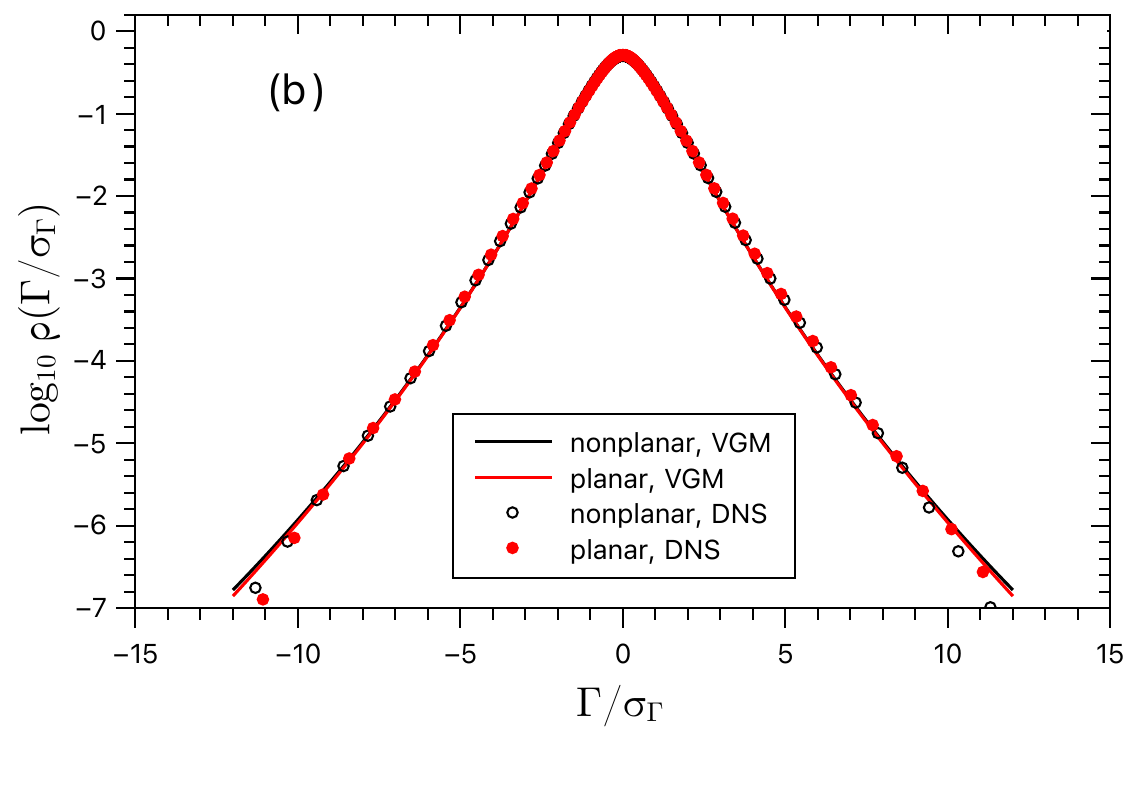}
\vspace{-0.5cm}
\caption{Comparison of cPDFs from planar and nonplanar VGM with DNS data (Ref.~\cite{Iyer_etal2}). Filled/empty symbols: DNS for planar/nonplanar contours of equal minimal area; solid lines: analytic model. (a) Nonplanar cPDFs standardized; planar circulations normalized by corresponding nonplanar standard deviations. (b) All cPDFs standardized.
}
\label{}
\end{figure}

Next, in Fig.~2b, we proceed to verify the area rule of circulation statistics, as evidenced from the collapse of cPDFs, when they are all provided in standardized form. As a robustness tests (not shown), we confirmed that cPDFs similarly collapse for planar rectangles with variable aspect ratios and identical minimal areas. In addition, the standardized cPDF for a smaller square of side $0.5R$ (still within the inertial range) shows a substantial deviation from those in Fig.~2b, as expected.

The agreement between the cPDFs predicted by the nonplanar VGM and the corresponding distributions measured in DNS is overall striking. We emphasize that no parameter tuning was performed. All model inputs are fixed by DNS and experimental data.

{\it{Conclusions.}} We have developed, under natural assumptions, a new formulation of the VGM for investigating the statistics of turbulent circulation fluctuations in nonplanar contours. The central outcome of the analysis is a precise yet phenomenologically grounded formulation of the cPDF as functional of the minimal surface bounded by the corresponding circulation contour. 

The long-conjectured connection between minimal surfaces and circulation statistics, first proposed by Migdal \cite{migdal2} as a general structural idea, and later critically reformulated through large scale numerical simulations \cite{Iyer_etal2} is here, for the first time, placed on a consistent theoretical foundation within the framework of the VGM. The comparison between the VGM-predicted and empirical cPDFs in Fig.~2 serves as an excellent benchmark for the model. It is worth noting that the VGM provides a complete description of cPDF shapes (encompassing both cores and tails) and elucidates the origin of circulation intermittency as a phenomenon intrinsically linked to the structure of the energy dissipation field.

A promising direction for future work is the refinement of modeled cPDFs through optimal surface corrections obtained from the solutions (\ref{expansion}) of Eq.~(\ref{sp}), together with surface fluctuations expected to be relevant for contours with linear dimensions near the lower end of the inertial range. Also, further improvement toward a slightly more accurate representation of the cPDF tails is likely to involve ensembles of surface vortex densities with explicit upper bounds, as previously examined in the context of the planar VGM \cite{bounded_measures, mori_pereira}.

The essential phenomenological content of the nonplanar VGM lies in the statistical organization of elementary vortices through their intermittent spatial distribution and circulation correlations in three dimensions. This insight opens new avenues for addressing problems across a broad range of systems where vortical structures are crucial, such as the sling effect in particle-laden flows \cite{falkovich_etal}, wall-bounded turbulence \cite{duan_etal}, and Rayleigh–Bénard convection \cite{schumacher}.

{\it{Acknowledgements.}}
The author gratefully acknowledges the warm hospitality of Katepalli Sreenivasan at the NYU, where this work was carried out. Enlightening discussions with Katepalli Sreenivasan and Sachin Bharadwaj are appreciated.~The DNS data was kindly shared by Kartik Iyer.~This work has been partially supported by the National Council for Scientific and Technological Development (CNPq).


\clearpage


\onecolumngrid
\appendix

\newcounter{ssec}
\renewcommand\thessec{S\arabic{ssec}}

\makeatletter
\@addtoreset{equation}{ssec}
\@addtoreset{figure}{ssec}
\@addtoreset{table}{ssec}
\makeatother

\renewcommand\theequation{\thessec.\arabic{equation}}
\renewcommand\thefigure{\thessec.\arabic{figure}}
\renewcommand\thetable{\thessec.\arabic{table}}

\newcommand{\suppsection}[1]{%
  \refstepcounter{ssec}%
  \section*{\thessec.\; #1}%
}

\section*{\large\underline{Supplemental Material}}

{\centerline{\large{\bf{Optimal Surfaces for Turbulent Circulation Statistics}}}}
\vspace{0.4cm}

{\centerline{L. Moriconi}}
{\centerline{\it{Instituto de F\'\i sica, Universidade Federal do Rio de Janeiro,}}}\
{\centerline{\it{C.P. 68528, 21945-970, Rio de Janeiro, RJ, Brazil and}}}
{\centerline{\it{Department of Mechanical and Aerospace Engineering,}}}
{\centerline{\it{New York University, New York, 11201, USA}}}

\suppsection{Det$[K]$ at Leading Order}

While a comprehensive review of heat-kernel applications can be found in \cite{vassilevich}, we provide a heuristic and accessible account of the main ideas relevant to the evaluation of the determinant in Eq.~(\ref{FzDet}). 
Consider, to start, the complete set of real and normalized eigenfunctions $\phi_n(x)$ of the operator $K$, with eigenvalues $\lambda_n$, which appear in the definition of the action (\ref{actionS}), that is,

\be
\int_\Sigma d^2x' \sqrt{g(x')} K(x,x'|z) \phi_n(x') = \lambda_n \phi(x) \ . \ 
\ee
An arbitrary field $\phi(x)$ can be expanded in that eigenfunction basis as $\phi(x) = \sum_n a_n \phi_n(x)$. Taking for granted that $K$ is a positive definite operator (the action is never negative), the functional integral in (\ref{f_energy}) is, in virtue of (\ref{actionS}),
\be
\int D[\phi] \exp\{-S[\phi,z] \} \propto \int \prod_n da_n \exp [ -\lambda_n\phi_n^2/2 ]
\propto \prod_n \lambda_n^{-1/2} \propto \{ {\hbox{Det}}[K] \}^{-1/2} \ . \  \label{detK}
\ee
At this point, we have just proved the first equality in (\ref{FzDet}). To proceed on, take, worrying with possible integral divergencies,
\be
{\mathcal{I}}[K] \equiv - \int_a^b \frac{dt}{t} {\hbox{Tr}} \{\exp [ -t K ] \} = - \sum_n \int_a^b \frac{dt}{t} \exp [ -t \lambda_n ] \ , \ \label{IK}
\ee
where $a,b>0$ with $b \gg a$ are regularizing cutoffs. We obtain
\be
\frac{d}{d\lambda_n} {\mathcal{I}}[K] =  \int_a^b dt \exp [ -t \lambda_n ] =  \frac{1}{\lambda_n} [\exp(-\lambda_n a) - \exp(-\lambda_n b)] \ , \
\ee
so that
\be
{\mathcal{I}}[K] \approx \sum_n [A(\lambda_n) + B(\lambda_n)\ln \lambda_n] \ , \ \label{IK2} 
\ee
where the above coefficients are (i) $A(\lambda_n),B(\lambda_n) \approx 0$ for $\lambda_n \gg 1/a$, (ii) $A(\lambda_n) \approx {\hbox{const}}, B(\lambda_n) \approx 1$ for $1/b \ll \lambda_n \ll 1/a$, and (iii) $A(\lambda_n) \approx \ln (b/a), B(\lambda_n) \approx 0$ for $\lambda_n \ll 1/b$. In the simplest case of a planar surface, $K$ reduces to the positive Laplacian $-\partial^2$, and the parameters $a$ and $b$ can be identified with the inverse squared ultraviolet and infrared wavenumber cutoffs, respectively: $a \sim \eta^2$ and $b \sim L^2$. For a curved surface, the ultraviolet cutoff remains essentially unchanged, as dictated by the short-distance behavior of (\ref{phi-phi}). The infrared cutoff, however, may vary since it is sensitive to the details of the circulation contour, though it should still scale with the integral length scale of the flow, yielding $b \sim L^2$. The second equality in (\ref{FzDet}) follows, now, as a consequence of (\ref{detK}), (\ref{IK}), and (\ref{IK2}).

The small $t$ asymptotic limit of the heat-kernel operator $K_t \equiv \exp [ -t K ]$ leads, ultimately, to (\ref{Fz}). In fact, $K_t$ satisfies the ``evolution" equation
\be
(\partial_t + K) K_t = 0 \ , \ \label{evol_eq}
\ee
with the initial condition $K_0 = \mathds1$, the identity operator in curved two-dimensional space. 
Eq. (\ref{evol_eq}) is formally identical, for small $t$, to the diffusion equation, with can be readily
solved to give
\be
K_t \approx \frac{1}{4 \pi t} \mathds1 \ . \ 
\ee
Substituting the above relation in (\ref{FzDet}), and recalling that ${\hbox{Tr}}[\mathds1] = \int_\Sigma d^2x \sqrt{g(x)}$, we obtain the regularized ultraviolet contribution to $F[z]$, Eq. (\ref{Fz}).

\suppsection{Numerical Evaluation of the Fluctuating Variances $\sigma^2[\xi]$}

We are interested to generate an ensemble of the random numerical variances (\ref{Gamma2}),
\be
\sigma^2[\xi] = \int_\Sigma d^2 x \int_\Sigma d^2 x'  \sqrt{g(x)} \sqrt{g(x)} \xi(x,z) \xi(x',z) \Delta(x,x') \ , \ \label{sigmav}
\ee
for a given minimal surface $z=z_0(x)$. For a square lattice with grid spacing $\eta = 1$, define the discrete positions $x_{ij} = (i,j)$, with $i$ and $j$ integers. All we have to do is to find the lattice kernel $G^{-1}(x_{ij},x_{kl})$ whose inverse gives the scalar correlation function (\ref{phi-phi}), that is, $G(x_{ij},x_{kl}) = \langle \phi(x_{ij}) \phi(x_{kl}) \rangle$. The kernel $G^{-1}(x_{ij},x_{kl})$ is then used to generate random realizations of $\phi(x_{ij})$ from the multivariate Gaussian distribution, 
\be
\rho[\{\phi(i,j)\}] \propto \exp \left ( -\frac{1}{2} \sum_i \sum_j \sum_k \sum_j
\phi(x_{ij},x_{kl}) G^{-1}(x_{ij},x_{kl}) \phi(x_{kl})  \right ) \ . \ \label{dist}
\ee
Taking $\xi_{ij} = \exp{ [\gamma \phi(x_{ij})/2]}$, the variance (\ref{sigmav}) is approximated by the Riemann sum,
\be
\sigma^2[\xi] = \sum_{i}\sum_{j}\sum_{k}\sum_{l} 
\sqrt{\vphantom{g(x_{kl})}g(x_{ij})} 
\sqrt{\vphantom{g(x_{ij})}g(x_{kl})}
\xi_{ij}\,\xi_{kl},\Delta(x_{ij},x_{kl}) \ . \ 
\label{sigmav2}
\ee
On an $M \times M$ lattice, we introduce the flattened representation of the kernel, 
\[
G_f(s,s') \equiv G(x_{ij},x_{kl}),
\]
together with the scalar fields 
\[
\phi_f(s) \equiv \phi(x_{ij}),
\]
where the single index $s \in \{1,\dots,M^2\}$ is defined via the mapping
\be
(i,j) \rightarrow s = i + M(j-1) \ , \ 
\ee
which can be inverted as
\bea
&&i = (s-i)\bmod{M} + 1  \ , \ \nonumber \\
&&j = \frac{s-i}{M} + 1 \ . \ 
\eea
Let $\{ \psi_n(s) \}$ denote the complete set of real eigenvectors of 
$G_f(s,s')$ with eigenvalues $\{ \zeta_n \}$,  
\be
\sum_{s'} G_f(s,s') \psi_n(s') = \zeta_n \psi_n(s) \, .
\ee
It follows from (\ref{dist}) that a random realization of the scalar field can be expressed as
\be
\phi_f(s) = \sum_n a_n \psi_n(s) \, ,
\ee
where $a_n \sim \mathcal{N}(0,\zeta_n)$. To match the resolution of the circular contour of Fig.~1, introduced in the DNS study of Ref.~\cite{Iyer_etal22}, we employ a lattice with linear size $M=150$. Since the investigated minimal surface is already well resolved at this resolution, an ensemble of $N = 10^4$ significantly faster yet reliable evaluations of the random variances (\ref{sigmav2}) were carried out on the decimated lattice of size $M/3 \times M/3$.
\\

All the codes used in this work (minimal surface generation, kernel inversion, random variances, cPDFs) were created by the author and are available upon request.


\begin{references}
\bibitem{comment} The more intermittent circulation field carried by the elementary vortex structures, closely related to $\tilde \Gamma$, is discussed in detail in \cite{mori_pereira2}.
\end{references}
\end{document}